\documentclass[doublecol]{epl2} 
\usepackage[colorlinks=true,linkcolor=blue,citecolor=red]{hyperref}
\usepackage{amsmath}
\usepackage{xcolor}
\usepackage{amssymb}
\usepackage[normalem]{ulem}

\newcommand{\REM}[1]{}
\newcommand{\bu}{\boldsymbol{u}}
\newcommand{\bx}{\boldsymbol{x}}
\newcommand{\br}{\boldsymbol{r}}
\newcommand{\bk}{\boldsymbol{k}}
\newcommand{\dd}{{\rm d}}
\newcommand{\bnabla}{\boldsymbol{\nabla}}
\newcommand{\filter}[2][K]{\overline{#2}_{#1}}
\title{Bridging Filtering and Point-Splitting Approaches for \\Variable-Density Flows}
\shorttitle{Filtering and Point-Splitting Approaches} 
\author{Hridey Narula\inst{1} \thanks{hrideynarula@tifrh.res.in} \and Prasad Perlekar \inst{1} \thanks{perlekar@tifrh.res.in}}
\shortauthor{Narula \textit{and} Perlekar}

\institute{                    
  \inst{1} Tata Institute of Fundamental Research, Hyderabad 500046, Telangana, India\\
}
\pacs{nn.mm.xx}{First pacs description}
\pacs{nn.mm.xx}{Second pacs description}
\abstract{Energy transfer in turbulent flows is typically described either through correlation functions, via the Kármán–Howarth–Monin relation, or through a scale-by-scale budget of the filtered energy~\cite{Frisch_1995}. For constant-density homogeneous and isotropic turbulence, the equivalence between these two descriptions is well understood.
In compressible turbulence, however, several definitions of filtered energy exist, and for most definitions the associated formulation in terms of correlation functions is unclear.
We develop a general empirical framework, supported by theoretical arguments and numerical simulations, to determine the multipoint correlation functions corresponding to any filtered energy.
We then show that the Favre filtered energy -- defined
as the ratio of the squared filtered momentum to the filtered density -- corresponds to an infinite series of multipoint correlation functions.
This is achieved by expanding the Favre velocity as a power series in local density fluctuations.
The expansion reveals the contributions of the subgrid-scale fluctuations of velocity and density to the Favre velocity.
We verify the proposed expansion for the buoyancy and pressure contributions for three-dimensional buoyancy-driven bubbly flows with a large density contrast ($10^2$) between the liquid and the bubble phase.
}
\begin{document}
\maketitle
\section{Introduction}
Turbulent flows inherently involve interactions across a wide range of scales. 
The energy injected at large scales is progressively transferred to smaller scales via nonlinear interactions present in the Navier-Stokes equations, before being dissipated by viscosity~\cite{Alexakis_2018}.
The presence of such an energy cascade across scales is a key feature of incompressible turbulence.
Inter-scale energy transfers are typically studied using the scale-by-scale budget equation, which describe the evolution of the kinetic energy contained in large scales~\cite{Pope_2000}.
The scale-by-scale budget can be established in two different, albeit related, ways: (a) the filtering/coarse-graining approach~\cite{Frisch_1995,Leonard_1975,Germano_1992,Pope_2000,EYINK_2005}, and (b) the point-splitting approach using two-point correlation functions (the Kármán-Howarth-Monin (KHM) relation)~\cite{KarmanHowarth_1938,monin2007statistical,Frisch_1995,HILL_2002,Davidson_2015}.
Although both of these approaches, as well as their correspondence, are well established for uniform-density homogeneous and isotropic turbulence (HIT)~\cite{Frisch_1995,Eyink_2006}, their application to the variable-density case remains nontrivial.
Ambiguity arises because variable-density flows admit multiple definitions of the large-scale kinetic energy~\cite{Aluie2013}.
For instance, two such possible definitions in the context of filtering are: filtered kinetic energy $\mathcal{E}(K)$ at scale $K$ defined as (i) the product of filtered velocity and filtered momentum fields $\filter{\rho\bu}\cdot\filter{\bu}/2$~\cite{Graham_2010}, and as (ii) the square of filtered momentum divided by the filtered density field (the Favre definition) $\filter{\rho\bu}^2/2\filter{\rho}$~\cite{Favre_1965}.
Here, $\rho$ is the density field, $\bu$ is the velocity field, and the overbar indicates the filtering operation.
In principle, there are infinitely many such definitions.
Each such possible definition of the large-scale kinetic energy gives rise to a different scale-by-scale budget, thereby yielding potentially different physical interpretations~\cite{ZhaoAluie_2018,Eyink_2018,Narula_2026jfm}.
Regardless, previous studies have shown that the Favre definition is the most appropriate, since it guarantees that the large-scale dynamics are not contaminated by viscous dissipation~\cite{Aluie2013,ZhaoAluie_2018}, and preserves the pure injection nature of the forcing term~\cite{Aluie2013,Narula_2026jfm}.\newline 
For the point-splitting approach, while multiple KHM relations have been derived using different two-point correlators (based on quadratic functions of the filtered fields)~\cite{Clark_1995,FALKOVICH_FOUXON_OZ_2010,GaltierBanerjee2011,Wagner_Falkovich_Kritsuk_Norman_2012,BanerjeeKritsuk_2017,Hellinger_2020,Arun_Sameen_Srinivasan_Girimaji_2021,Narula_2026jfm,Fabien_2025}, there are no exact formulations for non-quadratic definitions of the large-scale energy (for example, the Favre definition).
The goal of this paper is to establish correlation functions for such non-quadratic cases.

The rest of the manuscript is organized as follows. We first show that angular averaging of a field can be interpreted as a low-pass filtering operation.
Next, we show that the angular-averaged multipoint correlation functions can approximate the moments of the filtered fields.
Using these ideas, we establish our main result -- the Favre filtered budget can be expressed using a series of multipoint correlation functions. This is achieved by expanding the density-weighted velocity field in powers of the local density fluctuations.
We numerically validate our results using existing datasets for HIT~\cite{Perlekar_2019}, and buoyancy-driven bubbly flows~\cite{PandeyMitraPerlekar2023} (see table~\ref{tab.1dns_details}). 
\begin{table}[!h]
\caption{Details of the DNS datasets used. In both cases, the Navier-Stokes (NS)  equations are solved in a $2\pi$ periodic cubic box discretised with $N$ equidistant points along each direction.}
\label{tab.1dns_details}
\begin{center}
\begin{tabular}{|l|p{7cm}|}
\hline
{\tt R1}  & 3$d$ uniform-density HIT data with constant energy injection rate in modes $k\in[1,2]$ with $\text{Re}_\lambda=96$ and $N=512$~\cite{Perlekar_2019}.\\\hline
{\tt R2} & 3$d$ incompressible, buoyancy-driven bubbly flows with density ratio $100$, Bond number ${\rm Bo}=1.78$, Galilei number ${\rm Ga}=1489$, $\text{Re}_\lambda=88$, volume fraction $3.2\%$, and bubble wavenumber $K_D\approx6$ with $N=504$~\cite{PandeyMitraPerlekar2023}.\\\hline 
\end{tabular}
\end{center}
\end{table}

\section{Filtering and Point-Splitting Approaches}
We start by briefly recalling the filtering and point-splitting approaches in the context of uniform-density flows ($\rho=\rho_0$), obeying the Navier-Stokes (NS) equations in a $2\pi$ periodic cubic box,
\begin{equation}\label{eq:nse_incomp}
\begin{aligned}
\rho_0 D_t\bu = -\bnabla P+ \mu\nabla^2\bu+\boldsymbol{F},\;\;
    \bnabla\cdot\bu = 0,
\end{aligned}    
\end{equation}
where $\bu$ is the velocity field, $D_t\equiv\partial_t+\bu\cdot\bnabla$ is the material derivative, $P$ is the pressure field, $\mu$ is the fluid viscosity, and $\boldsymbol{F}$ is a generic large-scale force.
\subsection{The Filtering Approach~\cite{Germano_1992, Frisch_1995,Pope_2000}}
The filtered velocity field $\filter{\bu}$ is obtained by convolving the velocity field $\bu$ with an isotropic low-pass filter $G_r(\bx)$ that suppresses fluctuations on scales smaller than $r \sim 1/K$.
The convolution is conveniently evaluated in Fourier space as
$\filter{\bu}={\rm IFT}[G_K(\bk)\widehat{\bu}(\bk)]$, where $G_K(\bk) = (2\pi)^3{\rm FT}[G_r]$ and $\widehat{\bu}(\bk)={\rm FT}[\bu]$~\cite{Pope_2000}, and ${\rm FT}$ and ${\rm IFT}$ denote forward and backward Fourier transforms.
The statistics of the filtered velocity field are robust with respect to the choice of the filtering kernel $G_r$, provided it is smooth and non-negative~\cite{Vreman_Geurts_Kuerten_1994,BORUE_ORSZAG_1998}.
The large-scale energy is defined as $\mathcal{E}(K)=\langle\filter{\bu}^2\rangle/2$ (where $\langle\rangle$ denotes spatial averaging), and its time evolution is obtained using eq.~\eqref{eq:nse_incomp}~\cite{Frisch_1995,Pope_2000}.

\subsection{The Point-Splitting Approach~\cite{Frisch_1995,Davidson_2015}}
The large-scale energy is defined as the velocity-velocity correlator $\mathcal{R}(\br)=\langle\bu(\bx)\cdot\bu(\bx+\br)\rangle/2$, which, under certain constraints, can be heuristically interpreted as the energy contribution from eddies of size $\ge r$ oriented along the direction $\hat{\br}$~\cite{Davidson_2005}.
Therefore, the isotropic sector of this correlation function $\mathcal{R}(r)$ is a surrogate for the energy contained in scales $\ge r$.
The corresponding time evolution equation for $\mathcal{R}$ can be obtained from eq.~\eqref{eq:nse_incomp}, and is the KHM relation~\cite{Frisch_1995, Davidson_2015}.\newline 
The two approaches are related by noting that $\mathcal{R}(\br)=\int |\widehat{\bu}(\boldsymbol{\bk})|^2 {\rm e}^{{\rm i} \bk \cdot \br} \dd^3 \bk/2$
by the Wiener-Khinchin theorem~\cite{KARMAN_1951,Alexakis_2018,McComb_2014,McComb_2015,Hamba_2022,Mccomb_2024,kubo1991statistical}.
The isotropic sector of $\mathcal{R}(\br)$ is related to the shell-averaged energy spectrum $E(k)=\sum_{k-1/2<q\leq k+1/2}|\widehat{\bu}(\boldsymbol{q})|^2$ as $\mathcal{R}(r) = \int \dd k\,E(k) \sin(kr)/kr$~\cite{batchelor1953theory,monin2007statistical,Frisch_1995,Davidson_2015}.
Interpreting the sinc function as a low-pass filter that suppresses contributions from $k\gtrsim1/r$ leads to the realization $\mathcal{R}(r)\approx\mathcal{E}(K)$ (calculated using a sharp spectral filter with $K\sim1/r$).
In the following section we show, using theoretical arguments and numerical validation, that angular averaging and low-pass filtering are qualitatively the same in dimensions $d>1$.
\section{Angular Averaging as Filtering}
\begin{figure*}[t]
\centering
\includegraphics[width=0.95\linewidth]{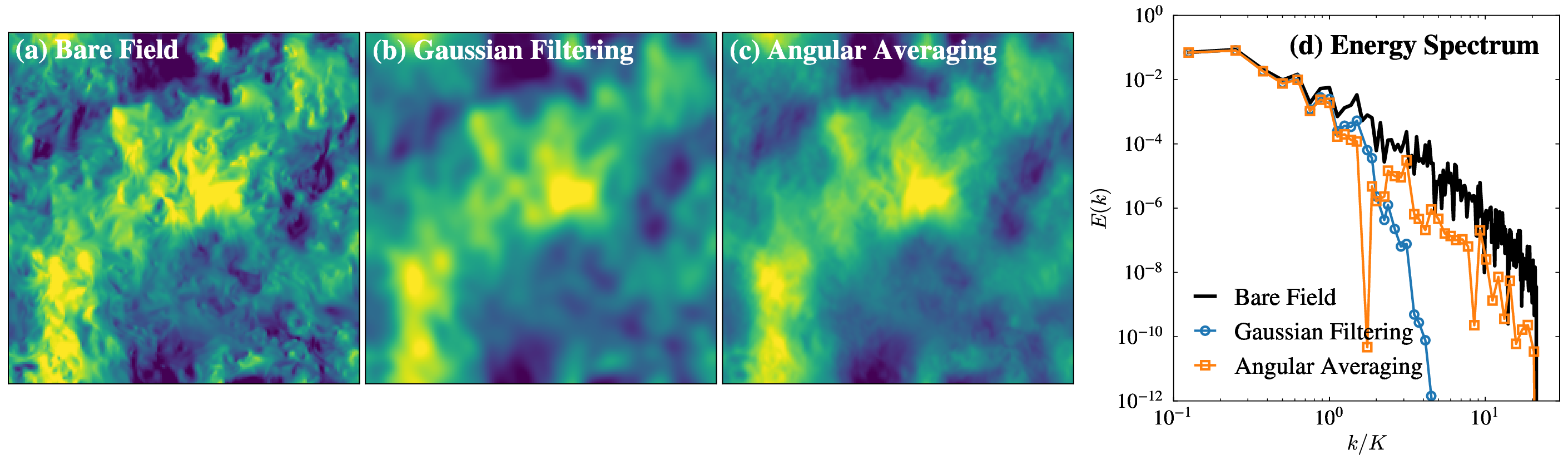}
\caption{Comparison of (a) the bare field $w(\bx)$, (b) the Gaussian filtered field $\filter{w}(\bx)$, and (c) the angular-averaged field $\filter{w}^\delta=\int\dd\theta w(\bx+\br)/2\pi$ for a 2$d$ slice from 3$d$ HIT DNS data in run {\tt R1}.
The filtering scale is $K=8$ and lies in the inertial range. We take $r=\sqrt{2}/K$ following ref.~\cite{Hellinger_2021b}.
Shell-averaged energy spectrum $E(k)$ (d) for $w$, $\filter{w}$ and $\filter{w}^\delta$ fields.
}\label{fig1:filtered_fields}
\end{figure*}
Consider a smooth $2\pi$-periodic field $\chi(\bx)$ in $d=3$.
Performing angular averaging over a length-scale $r$, we get 
\begin{equation}\label{eq:filter_diracdelta}
\begin{aligned}
    \int \frac{\dd \Omega}{4\pi} \chi(\bx+\br) &= \frac{1}{4\pi r^2}\int \dd^3 \br'\chi(\bx+\br')\delta(r'-r),\\
     &=\int \dd^3\bk \widehat{\chi}(\bk) \frac{\sin kr}{kr} {\rm e}^{{\rm i}\bk\cdot\bx} \equiv \filter{\chi}^\delta(\bx),
\end{aligned}
\end{equation}
where $\br'$ is a dummy integral variable, $r'=|\br'|$, and $\Omega$ is the solid angle associated with $\br'$.
In the second line above, we interpret $\sin(kr)/kr$ as the filter in Fourier space. The corresponding real-space filter is the radial delta function $G^\delta_r(r')=\delta(r'-r)/4\pi r^2$ and the filtered field is $\filter{\chi}^\delta$.
$G^\delta_r$, although not smooth, is non-negative, normalized, and has its support on the shell of radius $r$ centered at $\bx$.
Similarly, in $d$ dimensions the Fourier transform of the radial delta function~\cite{Vembu_1961} gives the equivalent filter (with $K\sim1/r$),
\begin{equation}\label{eq:filter_diracdelta_d}
    G_{K}^\delta (k) \equiv\frac{ (2\pi)^d}{S_d}\, {\rm{FT}}[\delta(r'-r)] = \frac{\Gamma(d/2)}{2^{1-d/2}}\frac{J_{d/2-1}(kr)}{(kr)^{d/2-1}},
\end{equation}
where $S_d$ is the surface area of the $d$-sphere, $J_b$ is the Bessel function of the first kind of order $b$ and $\Gamma$ is the gamma function.
Note that $G_K^\delta(k)\to1$ for $k\ll K$, while for $k\gg K$, $G_K^\delta(k)\sim (k/K)^{(1-d)/2}$ and hence can be  interpreted as a low-pass filter for $d>1$.

We now qualitatively verify this equivalence of angular averaging and filtering by considering a two-dimensional slice of the vertical component $w$ of a turbulent velocity field (run {\tt R1} in table~\ref{tab.1dns_details}). 
In fig.~\ref{fig1:filtered_fields}, we show that the filtered field $\filter{w}={\rm IFT}[G_K\widehat{w}]$ obtained by applying a Gaussian kernel $G_K(k)=\exp(-\pi^2k^2/24K^2)$~\cite{Pope_2000}, and the angular-averaged field $\int_0^{2\pi} {\rm d}\theta\, w({\bm x}+{\bm r})/2\pi$ ($\theta$ denotes the polar angle of $\br$) are qualitatively similar.
We have taken $K=8$, which lies in the inertial range.
The filter corresponding to angular averaging, $G_K^\delta(k)$, decays as $(k/K)^{-1/2}$ in $2d$, and therefore suppresses the small-scale features less than the Gaussian kernel, as is evident in the comparison of the shell-averaged spectrum $E(k)$ in fig.~\ref{fig1:filtered_fields}d.
\section{Angular-Averaged Multipoint Correlation Functions}
We now consider the general case for $d>1$. 
Let $C_p(r)$ be the $p$-point correlator of the field $\chi$ averaged over all possible orientations of the increment vector,
\begin{equation}\label{eq:corr_defn}
\begin{aligned}
    C_p(r) &= \prod_{i=1}^{p-1} \int\frac{\dd S_i}{S_d} \left\langle \chi(\bx)\ldots\chi(\bx+\br_{p-1}) \right \rangle,
\end{aligned}
\end{equation}
where $\dd S_i$ is the surface element for the $\br_i$ increment vector, $r_i=r$ for all $i$, and $S_d$ is the surface area of the $d$-dimensional sphere.
As discussed earlier, by eq.~\eqref{eq:filter_diracdelta_d}, we can write the correlator in terms of the filtered fields,
\begin{equation}
    C_p(r) =  \langle \chi (\filter{\chi}^\delta)^{p-1}\rangle.
\end{equation}
$C_p$ is, therefore, the product of one bare field with $p-1$ filtered fields with filter $G_K^\delta$. 
We define $A_p\equiv\langle\chi\filter{\chi}^{p-1}\rangle$ as the equivalent quantity evaluated with any arbitrary positive filtering kernel.
We will now show that $A_p\approx\mathcal{M}_p$, where $\mathcal{M}_p\equiv\langle\filter{\chi}^{p}\rangle$ is the $p$-th moment of the filtered field.

We first note that $A_p=\mathcal{M}_p$ for $K=0$ and $K=\infty$.
In particular, for $K\to\infty$ ($r\to0$), the difference $\chi(\bx)-\filter{\chi}(\bx)=\int\dd^d\br'\,G_r(r') (\chi(\bx)-\chi(\bx+\br'))$ scales as $\nabla^2\chi K^{-2}$ (obtained by by expanding $\chi(\bx+\br')$ to second order in $\br'$~\cite{LeesAluie_2019}) for smooth $\chi$, therefore $A_p\approx\mathcal{M}_p$ for large $K$.

We now obtain a bound on the difference $|A_p-\mathcal{M}_p|$  for arbitrary $K$, starting with the triangle inequality~\cite{Steele_2004},
\begin{eqnarray}
    |A_p-\mathcal{M}_p| &=& |\langle (\chi-\filter{\chi})\, \filter{\chi}^{p-1}\rangle|\nonumber\leq  \langle |\chi-\filter{\chi}|\, |\filter{\chi}|^{p-1} \rangle.\nonumber
\end{eqnarray} 
Using H\"older's inequality, $\langle |fg|\rangle\leq \|f\|_m\|g\|_n$ with $1/m+1/n=1$, and where $\|.\|_m$ is the $L_m$ norm ($\|f\|_m\equiv\langle|f|^m\rangle^{1/m}$)~\cite{Steele_2004} with $m=p$ and $n=p/(p-1)$,
\begin{eqnarray}\label{eq:absolute_bound}
    |A_p-\mathcal{M}_p|&\leq& \|\chi-\filter{\chi}\|_p\,\|\filter{\chi}\|_p^{p-1}.
\end{eqnarray}
The first term on the right side above can be rewritten in terms of structure functions.
Defining $\Delta_{\br'}\chi\equiv\chi(\bx+\br')-\chi(\bx)$ and using Jensen's inequality
$|\chi-\filter{\chi}|^p\leq\int{\rm d}^d\br'\,|\Delta_{\br'}\chi|^pG_r(r')$\cite{Eyink_1995} (with $r\sim1/K)$,
\begin{equation}\label{eq:scaling_bound}
	\begin{aligned}
		\|\chi-\filter{\chi}||_p&\equiv\langle|\chi-\filter{\chi}|^p\rangle^{1/p}\\
		&\leq\left\langle \int {\rm d}^d\br'|\Delta_{{\br'}}\chi|^p G_r(r') \right\rangle^{1/p}\\
		&\leq \gamma_1 \left(\int {\rm d}^d\br' |\br'|^{\sigma_pp} G_r(r')\right)^{1/p}\\
		&=\gamma_2 r^{\sigma_p} = \gamma_3 K^{-\sigma_p},
	\end{aligned}
\end{equation}
where we have assumed that $\chi$ is Besov regular with
exponent $\sigma_p>0$, that is, $\|\Delta_{\br'}\chi\|_p\leq\gamma_1^p|\br'|^{\sigma_p}$  for some constant $\gamma_1^p$\cite{Eyink_1995Besov}.
The constant $\gamma_2^p=\gamma_1^p\int{\rm d}^d\boldsymbol{s}\,G(\boldsymbol{s}) |\boldsymbol{s}|^{\sigma_p}$ depends on the shape of the (non-dimensional) filtering kernel $G(\boldsymbol{s})\equiv r^dG_r(r\boldsymbol{s})$ \cite{EYINK_2005} is finite if $G(\boldsymbol{s})$ decays sufficiently fast in the real space \cite{eyink_2008notes}.
Hence, the difference in \eqref{eq:absolute_bound} vanishes with increasing $K$ provided $\|\filter{\chi}\|_p^{p-1}$ grows slower than $K^{\sigma_p}$.
For even $p$ and non-vanishing $\mathcal{M}_p(K)$, we can put a bound on the relative error by noting that the second term in the relation \eqref{eq:absolute_bound} is $\mathcal{M}_p^{(p-1)/p}$.
Dividing both sides of \eqref{eq:absolute_bound} by $\mathcal{M}_p=\langle|\filter{\chi}|^p\rangle$, using the scaling in \eqref{eq:scaling_bound} and noting that $\|\filter{\chi}\|_p\ge\|\filter{\chi}\|_2$,
\begin{equation}\label{eq:upper_bound_diff}
    \frac{|A_p-\mathcal{M}_p|}{\mathcal{M}_p}\leq \frac{\|\chi-\filter{\chi}\|_p}{\|\filter{\chi}\|_p}\leq\gamma_3 \frac{K^{-\sigma_p}}{\|\filter{\chi}\|_2}.
\end{equation}
The upper bound in relation \eqref{eq:upper_bound_diff} decreases monotonically with increasing $K$ for filters which are monotonic functions of the cutoff $K$ (e.g., the Gaussian and sharp-spectral filters).
Thus, from \eqref{eq:absolute_bound} and \eqref{eq:upper_bound_diff} we can approximate $A_p=\langle\chi\filter{\chi}^{p-1}\rangle\approx\langle\filter{\chi}^p\rangle=\mathcal{M}_p$ for any positive filter $G_K$.
For the $G_K^\delta$ filter, corresponding to angular averaging,
\begin{equation}\label{eq:corr_self_filt_p}
\begin{aligned}
 C_p = \langle\chi(\filter{\chi}^\delta)^{p-1}\rangle\approx \langle (\filter{\chi}^\delta)^p\rangle.
\end{aligned}
\end{equation}
We remark here that the above correspondence between the correlators and filtered field holds only for the specific choice of the radial delta filter $G^\delta_K$.
If we assume that the statistics are robust with respect to the choice of the kernel~\cite{Vreman_Geurts_Kuerten_1994,BORUE_ORSZAG_1998}, we have:
\begin{equation}\label{eq:corr_filt_p}
    C_p(r)\approx\mathcal{M}_p(K),
\end{equation}
where $C_p$ is the angular-averaged $p$-point correlator of the field $\chi$, $\mathcal{M}_p$ is the $p$-th moment of the filtered field $\filter{\chi}$ (with an arbitrary positive filtering kernel), and $K\sim1/r$.
Note that replacing the radial delta filter $G_K^\delta$ in eq.~\eqref{eq:corr_self_filt_p} by a generic filter in eq.~\eqref{eq:corr_filt_p} is a delicate step, and one can construct synthetic fields for which such a replacement may not be mathematically justified.
For $p=2$, $|\mathcal{M}_2-C_2|=\int\dd^3\bk|G_K^\delta(k)-G^2_K(k)||\widehat{\chi}(\bk)|^2$ which can be large if $|\widehat{\chi}(\bk)|^2$ peaks at $k=k^*$ for which the difference $\varsigma=|G_K^\delta(k^*)-G_K^2(k^*)|$ is large.
Regardless, for the fields and filters we consider, we find empirically that approximation \eqref{eq:corr_filt_p} is valid.

\begin{figure}[!h]
\onefigure[scale=0.38]{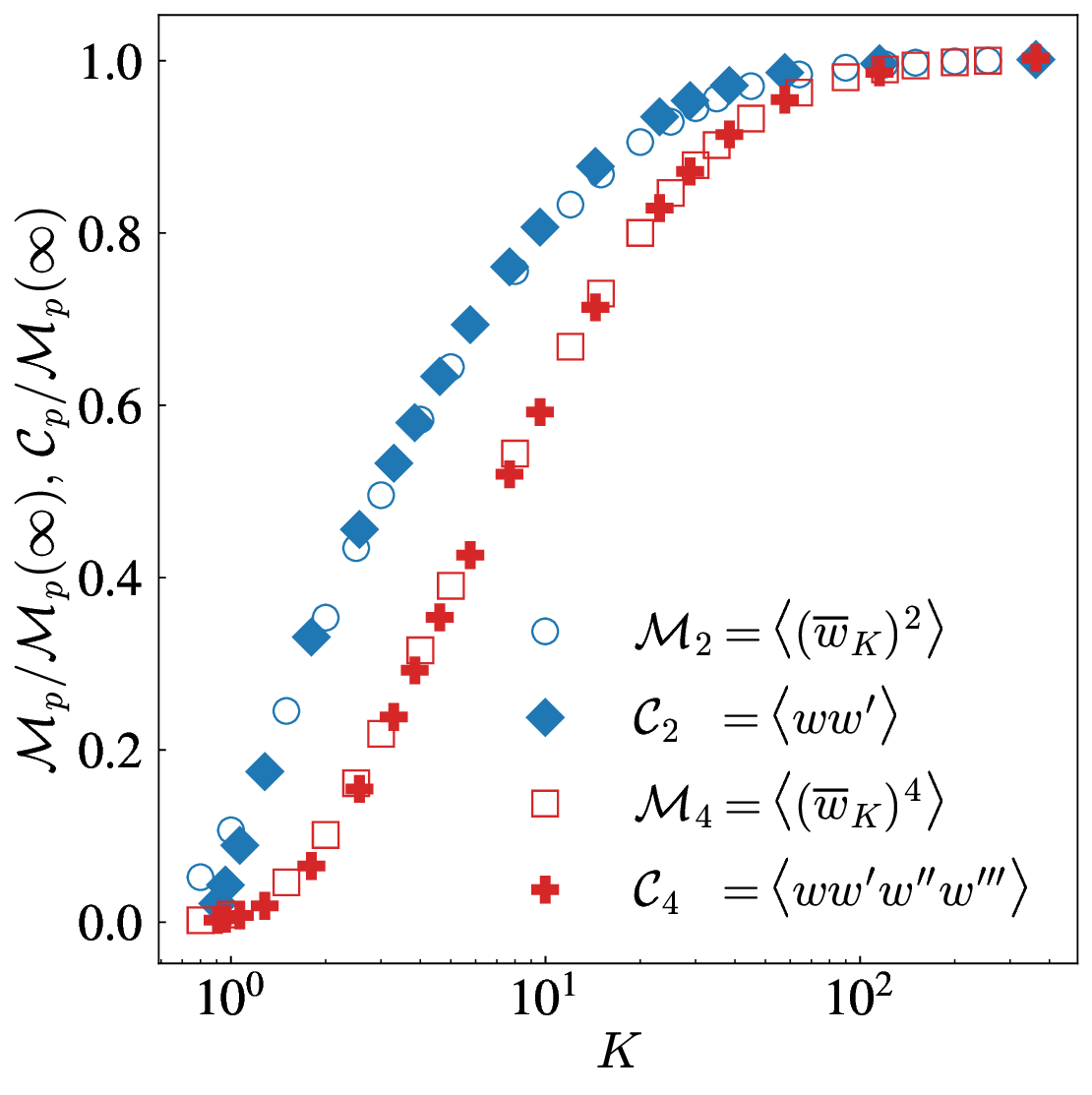} 
\caption{Comparison of the two-point $C_2$ and four-point $C_4$ correlators  with the second $\mathcal{M}_2$ and fourth $\mathcal{M}_4$ moments of the filtered fields for the 2$d$ slice of $w$ shown in fig.~\ref{fig1:filtered_fields} from 3$d$ uniform-density HIT DNS data (run {\tt R1}).
For correlation functions, we use the correspondence $K\cdot r=\sqrt{2}$~\cite{Hellinger_2021b}.}
\label{fig2:p2_4_compare_2dslice_3dHIT}
\end{figure}

In what follows, we always evaluate $\mathcal{M}_p$ with the Gaussian filter (for which $\varsigma\sim\mathcal{O}(10^{-1}$)) and $C_p$ by eq. \eqref{eq:corr_defn}.
We numerically verify eq.~\eqref{eq:corr_filt_p} for $p=2$ and $p=4$ using the same 2$d$ slice of the velocity field which is shown in fig.~\ref{fig1:filtered_fields}a.
Further, to plot correlators as functions of $K$, we use the empirical correspondence $K\cdot r = \sqrt{2}$~\cite{Hellinger_2021b}.
In fig.~\ref{fig2:p2_4_compare_2dslice_3dHIT} we show that $C_2\approx\mathcal{M}_2$ and $C_4\approx\mathcal{M}_4$ for all $K$. 
\subsection{Filtered energy and correlation functions in variable-density flows}
\begin{figure}[!h]
\onefigure[scale=0.38]{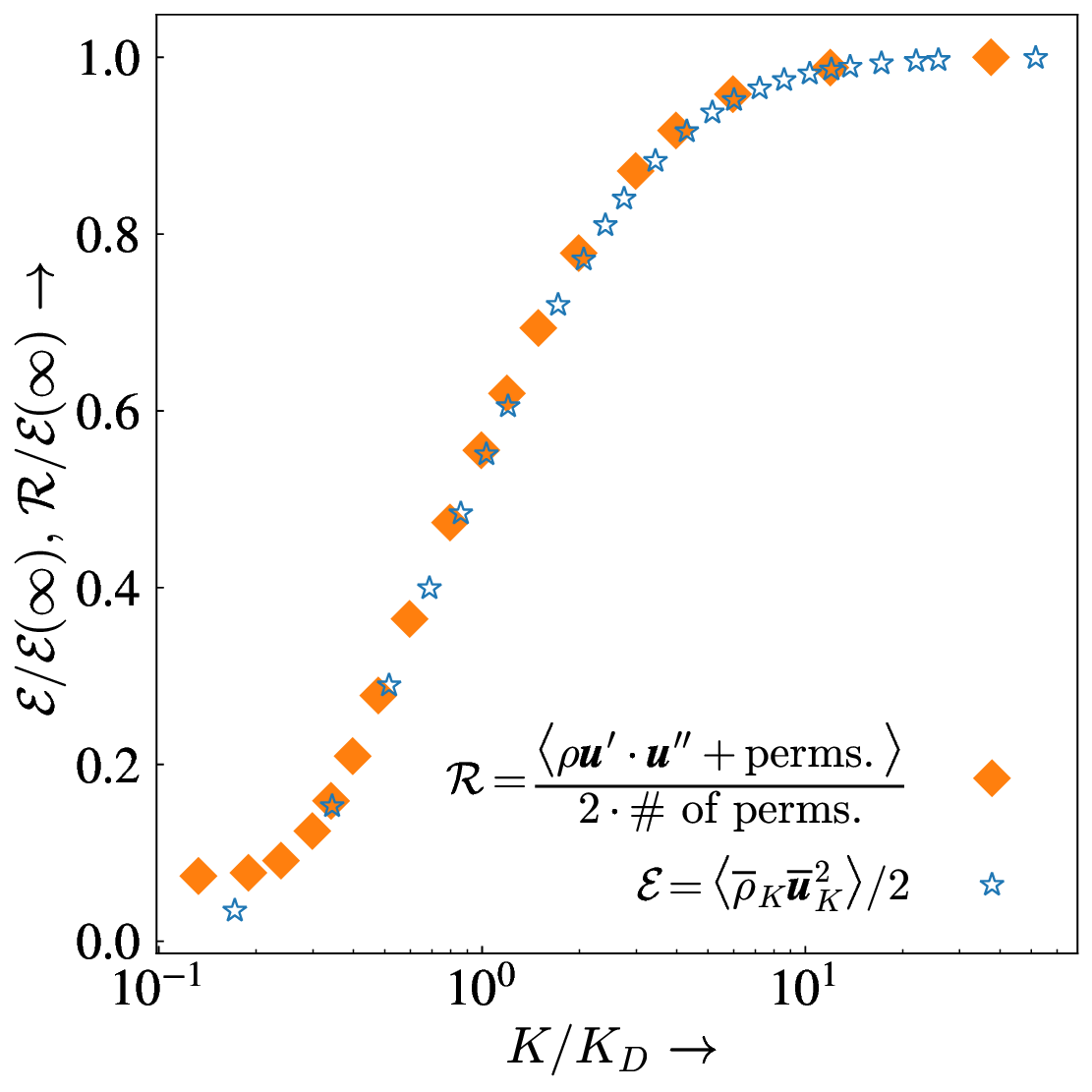} 
\caption{Comparison of the cubic large-scale energy as defined in table~\ref{tab.2energy_defns} for the $3d$ DNS data with bubbles (run {\tt R2}) with equations of motion \eqref{eq:nse_bub}. 
$K_D$ is the wavenumber corresponding to the bubble diameter and is roughly the injection scale.
For correlation functions, we use the correspondence $K\cdot r=\sqrt{2}$~\cite{Hellinger_2020}.}
\label{fig3:energy_f3}
\end{figure}

Consider a fluid with density field $\rho$ and velocity field $\bu$. 
The filtered energy for such a system is not unique, we list some possible definitions and the corresponding correlation functions in table~\ref{tab.2energy_defns}. 
For the first two definitions, the equivalence between the filtered energy (which are quadratic in the filtered fields), and the corresponding two-point correlation functions is well-established~\cite{Wang_2013,Hellinger_2021a,Hellinger_2021b,Narula_2026jfm}. 
In contrast, the third definition is cubic in the filtered fields.
Building on the arguments presented above (which are only valid for a single field), we propose ${\mathcal R(r)}$ to be a (symmetric) three-point correlator of $\rho$ and $\bu$ (see table \ref{tab.2energy_defns}).
We numerically verify the correspondence between ${\mathcal E}(K)$ and ${\mathcal R}(r)$ for the third (cubic) definition in fig.~\ref{fig3:energy_f3} using run {\tt R2} from table~\ref{tab.1dns_details}.
We observe good agreement between $\mathcal{E}(K)$ and $\mathcal{R}(r)$ with the largest discrepancy near $K\approx0$ where the approximation in \eqref{eq:corr_filt_p} has the largest error.
This concludes our validation for the equivalence between moments of filtered fields and multipoint correlation functions using eqs.~\eqref{eq:filter_diracdelta} and~\eqref{eq:corr_filt_p}. 
\begin{table}
\caption{A few definitions of filtered energy and correlation functions employed in the analysis of variable-density flows. Superscripts $'$ and $''$ denote evaluation of the fields at $\bx+\br_1$ and $\bx+\br_2$, respectively, with $r_1=r_2=r\sim1/K$.}
\label{tab.2energy_defns}
\begin{center}
\begin{tabular}{|c|c|}
\hline 
$2\cdot\mathcal{E}(K)$& $2\cdot\mathcal{R}(r)$ \\\hline
$\langle\filter{\sqrt{\rho}\bu}^2\rangle \text{\cite{Kida_1990}}$ & $\langle \sqrt{\rho}\bu\cdot\sqrt{\rho'}\bu' \rangle \text{\cite{Wang_2013}}$ \\
$\langle\filter{\rho\bu}\cdot\filter{\bu}\rangle \text{\cite{Graham_2010}}$ & $\langle\rho\bu\cdot\bu'+\rho'\bu'\cdot\bu\rangle/2\text{\cite{GaltierBanerjee2011}}$\\
$\langle\filter{\rho}\cdot\filter{\bu}^2\rangle \text{\cite{CHassaing1985}}$&$\langle\rho\bu'\cdot\bu''+\text{perms.}\rangle/\text{num. of perms.}$\\\hline 
\end{tabular}
\end{center}
\end{table}
However, as highlighted in the introduction, recent studies~\cite{Aluie2013,ZhaoAluie_2018,Eyink_2018,Narula_2026jfm} show that it is physically most appropriate to study the scale-by-scale budget using the Favre filtered energy ${\mathcal E}(K)=\langle\filter{\rho\bu}\cdot\filter{\rho\bu}/\filter{\rho}\rangle/2$, for which no obvious point-splitting approach can be obtained, because $\mathcal{E}(K)$ is a non-polynomial function of the filtered fields. 

This brings us to the main result of our manuscript.
In the next section, we show that the Favre filtered contributions can be written as the sum of an infinite series of multipoint correlation functions.
\section{A Point-Splitting Analogue to the Favre Filtered Budget}
Let $\rho$ be the density field and $\bu$ be the velocity field.
The Favre velocity $\widetilde{\bu}_K$ and the Favre filtered energy $\mathcal{E}(K)$ are then defined as~\cite{Favre_1965},
\begin{equation}\label{eq:favre_vel_defn}
    \begin{aligned}
    \widetilde{\bu}_K =\frac{\filter{\rho\bu}}{\filter{\rho}},\, \text{and},\;\:
    \mathcal{E}(K) = \frac{1}{2} \left\langle \widetilde{\bu}_K\cdot\filter{\rho\bu} \right\rangle.    
    \end{aligned}
\end{equation}
We can formally expand the above expressions as a polynomial series under the modest assumption that the density field is bounded from above.
Consider a reference density value $\rho_0$ that satisfies the following constraint: $|(\rho-\rho_0)/\rho_0|<1$ everywhere. The local fluctuations in the density field about $\rho_0$ are $\psi(\bx) = \rho(\bx)/\rho_0-1$, and we can write $\rho(\bx)=\rho_0(1+\psi(\bx))$ with $\psi$ being the ``small" parameter, that is, $\psi(\bx)<1$ for all $\bx$.
The Favre velocity $\widetilde{\bu}_K$ can then be expanded as,
\begin{align}\label{eq:favereu_expansion}
    \widetilde{\bu}_K &= (\filter{\bu}+\filter{\psi \bu}) (1-\filter{\psi}+\filter{\psi}^2+\mathcal{O}(\psi^3))\\
     &= \filter{\bu} + \filter{\tau}(\psi,\bu) -  \filter{\tau}(\psi,\bu)\filter{\psi} + \mathcal{O}(\psi^3),
\end{align}
where $\mathcal{O}(\psi^3)$ denotes third and higher order terms in $\psi$.
$\filter{\tau}(\psi,\,\bu)=\filter{\psi\bu}-\filter{\psi}\filter{\bu}$ is a second cumulant of the filtered fields and represents the subgrid-scale fluctuations between $\psi$ and $\bu$ fields~\cite{EYINK_2005,Eyink_2009,Eyink_2018,Capocci_2025}.
Eq.~\eqref{eq:favereu_expansion} can be interpreted as follows: the leading contribution to the Favre velocity comes from the filtered velocity $\filter{\bu}$.
The first-order correction arises because of the subgrid-scale fluctuations between $\psi$ and $\bu$.
Every successive correction is due to the interaction of the subgrid-scale fluctuations with the filtered  density-fluctuation field $\filter{\psi}$. Using the expansion of $\widetilde{\bu}_K$ in eq.~\eqref{eq:favre_vel_defn}, we obtain 
\begin{align}\label{eq:favre_energy_expansion}
    \mathcal{E}(K) = \frac{\rho_0}{2} \left\langle (\filter{\bu}+\filter{\tau}(\psi,\bu) +\ldots)\cdot(\filter{\bu}+\filter{\psi\bu}) \right \rangle.\nonumber
\end{align}
The leading order term ($\langle \rho_0 \filter{\bu}^2\rangle/2$ which corresponds to the Boussinesq approximation), can be expressed as the two-point correlator $\rho_0\langle \bu \cdot \bu' \rangle/2$. The subsequent $\mathcal{O}(\psi^m)$ terms in the expansion are composed of at most $m+2$ point correlators  (see eq.~\eqref{eq:corr_filt_p}). Therefore, the Favre filtered energy can be expressed as a series of correlation functions.
Note that the total quadratic contribution  $\langle\filter{\rho\bu}^2\rangle/2\rho_0$  corresponds to the two-point momentum-momentum correlator~\cite{FALKOVICH_FOUXON_OZ_2010,Wagner_Falkovich_Kritsuk_Norman_2012} which has a regularized KHM relation in the infinite Reynolds number limit~\cite{Eyink_2018}.
\section{Buoyancy-Driven Bubbly Flows}
\begin{figure*}[t]  
\centering
\includegraphics[scale=0.4]{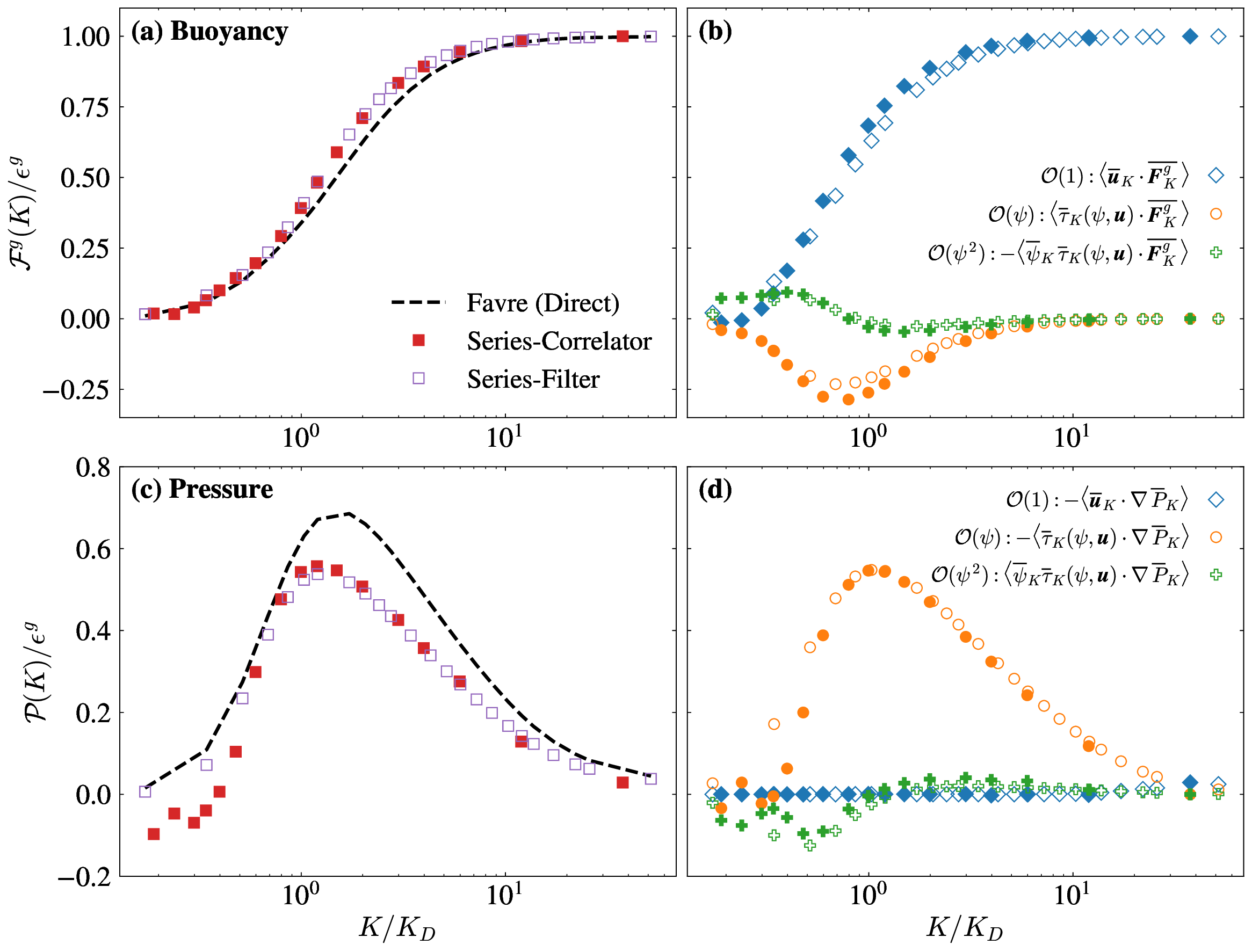}
\caption{\textbf{Top}: (a) Evaluation of the Favre filtered contribution for the buoyancy term alongside the series expansion up to second order (eq. \eqref{eq:buoy_expand}) using correlation functions (filled markers) and filtered fields (empty markers) and (b) contribution to the series expansion order-by-order.
\textbf{Bottom}: Same comparison for the pressure contribution (eq. \eqref{eq:pres_expand}).
Here, $\epsilon^g=\langle\bu\cdot\boldsymbol{F}^g\rangle$ is the total (bare) energy injection. For correlation functions, we have used the correspondence $K=\sqrt{3}/r$ \cite{Hellinger_2020}.
}\label{fig4:combine_buoyancy_baropycnal_term}
\end{figure*}
We now verify the series expansion in eq.~\eqref{eq:favereu_expansion} for the case of dilute (volume fraction $\approx3.2\%$) 3$d$ incompressible bubbly flows (run {\tt R2} in table~\ref{tab.1dns_details}).
The one-phase formulation of the Navier-Stokes equations for multiphase flows is~\cite{BUNNER_TRYGGVASON_2002}: 
\begin{eqnarray}\label{eq:nse_bub}
    \partial_t\rho\bu + \bnabla\cdot\rho\bu\bu &=& -\bnabla P+\mu\nabla^2\bu + \boldsymbol{F}^g+\boldsymbol{F}^\sigma,\\
    \partial_t\rho+\bnabla\cdot\rho\bu &=& 0;\;
    \bnabla\cdot\bu=0,
\end{eqnarray}    
where $P$ is the hydrodynamic pressure, $\mu$ is the (constant) viscosity, $\boldsymbol{F}^g=(\rho-\langle\rho\rangle)\boldsymbol{g}$ is the buoyancy force with gravitational acceleration $\boldsymbol{g}$ and $\boldsymbol{F}^\sigma=\sigma\kappa\boldsymbol{n}$ is the surface tension force where $\sigma$ is the surface tension coefficient, $\kappa$ is the local curvature of the interface and $\boldsymbol{n}$ is the normal vector to the interface.
The density in the bubble and the liquid phases is $\rho_B$ and $\rho_L$ respectively.
For the present study, we consider the case of large-density ratio $\rho_L/\rho_B=100$.
The details about numerical simulations and validation are discussed in refs. ~\cite{BUNNER_TRYGGVASON_2002,Pandey_Ramadugu_Perlekar_2020,Ramadugu_Pandey_Perlekar_2020,Aniszewski2021,PandeyMitraPerlekar2023}.
The Favre filtered budget can be obtained from eq.~\eqref{eq:nse_bub}, see, for example,~\cite{Aluie2013,PandeyMitraPerlekar2023,Narula_2026jfm},
\begin{equation}\label{eq:favre_budget}
    \partial_t\mathcal{E}(K) = \mathcal{N}(K)+\mathcal{P}(K)+\mathcal{D}(K)+\mathcal{F}^g(K)+\mathcal{F}^\sigma(K),
\end{equation}
where $\mathcal{N}(K)=\filter{\rho}\bnabla\widetilde{\bu}_K:(\widetilde{\bu\bu}_K-\widetilde{\bu}_K\widetilde{\bu}_K)$ is the nonlinear flux, $\mathcal{P}(K)=-\langle\widetilde{\bu}_K\cdot\bnabla\,\filter{P}\rangle$ is the pressure contribution, $\mathcal{D}(K)=\mu\langle\widetilde{\bu}_K\cdot\nabla^2\filter{\bu}\rangle$ is the viscous dissipation, $\mathcal{F}^g(K)=\langle\widetilde{\bu}_K\cdot\filter{\boldsymbol{F}^g}\rangle$ is the buoyancy injection and $\mathcal{F}^\sigma(K)=\langle\widetilde{\bu}_K\cdot\filter{\boldsymbol{F}^\sigma}\rangle$ is the transfer term due to surface tension.
In the statistically steady state ($\partial_t\mathcal{E}(K)\approx0$) we get a balance between the different terms.
We note here that $\mathcal{N},\mathcal{P}$ and $\mathcal{F}^\sigma,$ vanish as $K\to\infty$ and are thus flux-like terms, while $\mathcal{F}^g$ and $\mathcal{D}$ saturate in this limit and thus represent source/sink-like terms.
\newline 
Recently, it has been shown that the role of both the buoyancy and pressure terms is sensitive to the choice of definition of the large-scale energy in the scale-by-scale budget~\cite{Narula_2026jfm}.
In particular, only the Favre definition preserves the pure injection nature of the buoyancy contribution.
Hence, to test our series expansion, we evaluate the buoyancy and pressure contributions in the Favre budget~\eqref{eq:favre_budget} and the corresponding series expansion, in terms of the filtered fields (with the Gaussian filter) as well as correlation functions.
The series expansion is evaluated up to second order in density fluctuations $\psi$, and we take $\rho_0=(\rho_B+\rho_L)/2$.
The calculation of correlation functions is computationally expensive since we need to take into account all possible permutations of the fields.
We perform angular averaging over the 6 Cartesian directions ($\pm\hat{\bx},\pm\hat{\boldsymbol{y}},\,{\rm and} \pm\hat{\boldsymbol{z}}$) and $252^3$ centers for each increment vector $\br_i$. Following ref.~\cite{Hellinger_2020}, we use the correspondence $K=\sqrt{3}/r$ to plot all correlators as a function of $K$. 
\subsection{Buoyancy Contribution}
The buoyancy contribution is $\mathcal{F}^g(K)=\langle\widetilde{\bu}_K\cdot\filter{\boldsymbol{F}^g}\rangle$.
For bubbly flows, it injects energy at large-scales, corresponding roughly to the bubble diameter $D$~\cite{Pandey_Ramadugu_Perlekar_2020}.
We now express $\mathcal{F}^g(K)$ as a series expansion in $\psi$ using eq.~\eqref{eq:favereu_expansion}, 
\begin{equation}\label{eq:buoy_expand}
\begin{aligned}
    \mathcal{F}^g(K) &= \langle\filter{\bu}\cdot\filter{\boldsymbol{F}^g}\rangle +\langle\filter{\tau}(\psi, \bu)\cdot\filter{\boldsymbol{F}^g}\rangle -\\& \langle\filter{\psi}\filter{\tau}(\psi, \bu)\cdot\filter{\boldsymbol{F}^g}\rangle +\mathcal{O}({\psi}^3).
\end{aligned}    
\end{equation}
We evaluate terms up to second order in $\psi$, these involve two, three and four-point correlation functions.
We find excellent agreement between the exact evaluation of the buoyancy term and the series expansion~\eqref{eq:buoy_expand} (evaluated using both correlators and filters) as shown in fig.~\ref{fig4:combine_buoyancy_baropycnal_term}a,b.
\subsection{Pressure Contribution}
The pressure contribution is $\mathcal{P}(K)=-\langle\widetilde{\bu}_K\cdot\bnabla\,\filter{P}\rangle$.
We now again expand the Favre velocity to get the following series for $\mathcal{P}(K)$,
\begin{equation}\label{eq:pres_expand}
\begin{aligned}
    \mathcal{P}(K) &= -\langle\filter{\bu}\cdot\bnabla\,\filter{P}\rangle -\langle\filter{\tau}(\psi, \bu)\cdot\bnabla\,\filter{P}\rangle + \\&\langle\filter{\psi}\filter{\tau}(\psi, \bu)\cdot\bnabla\,\filter{P}\rangle+\mathcal{O}({\psi}^3).
\end{aligned}    
\end{equation}
In fig.~\ref{fig4:combine_buoyancy_baropycnal_term}c we show the qualitative agreement between the exact calculation of the Favre filtered pressure contribution and the series expansion in eq.~\eqref{eq:pres_expand}. 
Note that $\mathcal{P}(K)>0$, which, in our sign convention, indicates energy transfer from small to large scales~\cite{PandeyMitraPerlekar2023,Narula_2026jfm}.
Since the flow is incompressible, the zeroth order contribution $\langle\filter{\bu}\cdot\bnabla\,\filter{P}\rangle$ vanishes at all scales~\cite{Frisch_1995}, so that higher-order terms are required for a better agreement (we have verified that our results do not change when we average over independent snapshots).
In fig.~\ref{fig4:combine_buoyancy_baropycnal_term}d we show the agreement between the filtering and point-splitting approaches for the terms in the above expansion.
The largest difference between the two is for small $K$ (large-scales), where the approximation in eq.~\eqref{eq:corr_filt_p} has the largest error.
Regardless, the two approaches are in agreement for $K\gtrsim K_D$ where $K_D$ is the wavenumber corresponding to the bubble diameter $D$, and is roughly the energy injection scale.
The first order contribution is responsible for the inverse transfer, which is partially suppressed by the $\psi^2$ contribution.
\section{Conclusion}
Our study provides a unified way of looking at inter-scale energy transfers in generic variable-density flows.
The connection between the two-point velocity correlations and the filtered energy (square of the  filtered velocity fields) is well-known for uniform-density HIT~\cite{batchelor1953theory,Frisch_1995,KARMAN_1951,McComb_2015,Davidson_2015}. However, such a correspondence remains elusive for general variable-density flows. 
We address this gap by first showing that angular averaging is equivalent to low-pass filtering in $d>1$ dimensions, with the filtering kernel $G^\delta_K$,
allowing us to write the $p$-point angular-averaged correlator $C_p(r)=\langle\chi(\filter{\chi}^{\delta})^{p-1}\rangle$.
We then show that $\langle\chi\filter{\chi}^{p-1}\rangle\approx\langle\filter{\chi}^p\rangle$ for an arbitrary positive filter (see bounds~\eqref{eq:absolute_bound} and~\eqref{eq:upper_bound_diff}).
Combining these two ideas, we obtain the approximation $C_p\approx\mathcal{M}_p$ with $\mathcal{M}_p$ being the $p$-th moment of the filtered fields evaluated with an arbitrary positive filter.
Although this final step is empirical, it is supported by numerical evidence for the velocity fields considered here (obtained from DNS of HIT and bubbly flows).
This allows us to identify the point-splitting analogues for the terms in the Favre filtered energy budget, which has been shown to be the most appropriate choice to study energy transfers in variable-density flows. 
We show that the Favre velocity field can be expanded as a power series in the local density fluctuations and hence different terms in the Favre budget can be evaluated using an infinite series of multipoint correlation functions.
While the proposed series expansion only assumes that the density field is bounded from above, we expect the rate of convergence of the series to depend on the flow parameters (for example, the density ratio in two-phase flows and the Mach number in compressible flows).
Further, the results of ref.~\cite{Aluie2013} on the inviscid criteria are also valid for our series expansion, therefore, for fixed $K$, as $\mu\to0$, $\mathcal{D}(K)\to0$ order by order.
We believe that our study would motivate further investigations on energy transfer mechanisms using both correlation functions and filtered fields in variable density flows.
\acknowledgments 
The authors thank Vikash Pandey and Dhrubaditya Mitra for fruitful discussions.
The authors acknowledge support from the Department of Atomic Energy (DAE), India under Project Identification No. RTI 4007, and DST
(India) Projects No. MTR/2022/000867.


\begin{thebibliography}{10}
	\expandafter\ifx\csname url\endcsname\relax\def\url#1{\texttt{#1}}\fi
	
	\bibitem{Frisch_1995}
	\Name{Frisch U.} \Book{Turbulence: The Legacy of A. N. Kolmogorov} (Cambridge
	University Press) 1995.
	
	\bibitem{Alexakis_2018}
	\Name{Alexakis A.~E. \and Biferale L.} \REVIEW{Phys. Rep.}{767-769}{2018}{1}.
	
	\bibitem{Pope_2000}
	\Name{Pope S.~B.} \Book{Turbulent Flows} (Cambridge University Press) 2000.
	
	\bibitem{Leonard_1975}
	\Name{{Leonard} A.} \REVIEW{Adv. Geophys.}{18}{1975}{237}.
	
	\bibitem{Germano_1992}
	\Name{Germano M.} \REVIEW{J. Fluid Mech.}{238}{1992}{325}.
	
	\bibitem{EYINK_2005}
	\Name{Eyink G.~L.} \REVIEW{Physica D}{207}{2005}{91}.
	
	\bibitem{KarmanHowarth_1938}
	\Name{{de Karman} T. \and {Howarth} L.} \REVIEW{Proc. R. Soc. Lond.
		A}{164}{1938}{192}.
	
	\bibitem{monin2007statistical}
	\Name{Monin A. \and Yaglom A.} \Book{Statistical Fluid Mechanics: Mechanics of
		Turbulence} (Dover Publications) 2007.
	
	\bibitem{HILL_2002}
	\Name{Hill R.~J.} \REVIEW{J. Fluid Mech.}{468}{2002}{317}.
	
	\bibitem{Davidson_2015}
	\Name{Davidson P.} \Book{{Turbulence: An Introduction for Scientists and
			Engineers}} (Oxford University Press) 2015.
	
	\bibitem{Eyink_2006}
	\Name{Eyink G.~L. \and Sreenivasan K.~R.} \REVIEW{Rev. Mod.
		Phys.}{78}{2006}{87}.
	
	\bibitem{Aluie2013}
	\Name{Aluie H.} \REVIEW{Physica D}{247}{2013}{54}.
	
	\bibitem{Graham_2010}
	\Name{Graham J.~P., Cameron R. \and Schüssler M.} \REVIEW{Astrophys.
		J.}{714}{2010}{1606}.
	
	\bibitem{Favre_1965}
		\Name{Favre A.~J.} \Book{The equations of compressible turbulent gases,}
	\newblock Annual Summary Report No. 1 (1965).
	\bibitem{ZhaoAluie_2018}
	\Name{Zhao D. \and Aluie H.} \REVIEW{Phys. Rev. Fluids}{3}{2018}{054603}.
	
	\bibitem{Eyink_2018}
	\Name{Eyink G.~L. \and Drivas T.~D.} \REVIEW{Phys. Rev. X}{8}{2018}{011022}.
	
	\bibitem{Narula_2026jfm}
	\Name{Narula H., Pandey V., Mitra D. \and Perlekar P.} \REVIEW{J. Fluid
		Mech.}{1032}{2026}{A12}.
	
	\bibitem{Clark_1995}
	\Name{Clark T.~T. \and Spitz P.~B.} Tech. Rep. Los Alamos National Lab., NM
	(United States) (06 1995).
	
	\bibitem{FALKOVICH_FOUXON_OZ_2010}
	\Name{Falkovich G., Fouxon I. \and Oz Y.} \REVIEW{J. Fluid
		Mech.}{644}{2010}{465}.
	
	\bibitem{GaltierBanerjee2011}
	\Name{Galtier S. \and Banerjee S.} \REVIEW{Phys. Rev.
		Lett.}{107}{2011}{134501}.
	
	\bibitem{Wagner_Falkovich_Kritsuk_Norman_2012}
	\Name{Wagner R., Falkovich G., Kritsuk A.~G. \and Norman M.~L.} \REVIEW{J.
		Fluid Mech.}{713}{2012}{482}.
	
	\bibitem{BanerjeeKritsuk_2017}
	\Name{Banerjee S. \and Kritsuk A.~G.} \REVIEW{Phys. Rev. E}{96}{2017}{053116}.
	
	\bibitem{Hellinger_2020}
	\Name{Hellinger P., Verdini A., Landi S., Franci L., Papini E. \and Matteini
		L.} \Book{On cascade of kinetic energy in compressible hydrodynamic
		turbulence,}
	\newblock arXiv:2004.02726 [physics.flu-dyn] (2020).
	\bibitem{Arun_Sameen_Srinivasan_Girimaji_2021}
	\Name{Arun S., Sameen A., Srinivasan B. \and Girimaji S.~S.} \REVIEW{J. Fluid
		Mech.}{920}{2021}{A31}.
	
	\bibitem{Fabien_2025}
	\Name{Thiesset F. \and Vahé J.} \REVIEW{J. Fluid Mech.}{1025}{2025}{A63}.
	
	\bibitem{Perlekar_2019}
	\Name{Perlekar P.} \REVIEW{J. Fluid Mech.}{873}{2019}{459}.
	
	\bibitem{PandeyMitraPerlekar2023}
	\Name{Pandey V., Mitra D. \and Perlekar P.} \REVIEW{Phys. Rev.
		Lett.}{131}{2023}{114002}.
	
	\bibitem{Vreman_Geurts_Kuerten_1994}
	\Name{Vreman B., Geurts B. \and Kuerten H.} \REVIEW{J. Fluid
		Mech.}{278}{1994}{351}.
	
	\bibitem{BORUE_ORSZAG_1998}
	\Name{Borue V. \and Orszag S.~A.} \REVIEW{J. Fluid Mech.}{366}{1998}{1}.
	
	\bibitem{Davidson_2005}
	\Name{Davidson P.~A. \and Pearson B.~R.} \REVIEW{Phys. Rev.
		Lett.}{95}{2005}{214501}.
	
	\bibitem{KARMAN_1951}
	\Name{von K\'arm\'an T. \and Lin C.~C.} \REVIEW{Rev. Mod.
		Phys.}{21}{1949}{516}.
	
	\bibitem{McComb_2014}
	\Name{McComb W.~D., Yoffe S.~R., Linkmann M.~F. \and Berera A.} \REVIEW{Phys.
		Rev. E}{90}{2014}{053010}.
	
	\bibitem{McComb_2015}
	\Name{McComb W.~D., Berera A., Yoffe S.~R. \and Linkmann M.~F.} \REVIEW{Phys.
		Rev. E}{91}{2015}{043013}.
	
	\bibitem{Hamba_2022}
	\Name{Hamba F.} \REVIEW{J. Fluid Mech.}{931}{2022}{A34}.
	
	\bibitem{Mccomb_2024}
	\Name{McComb D.} \Book{What is isotropic turbulence and why is it important?,}
	\newblock arXiv:2403.13962 [math-ph] (2024).
	
	\bibitem{kubo1991statistical}
	\Name{Kubo R., Toda M. \and Hashitsume N.} \Book{Statistical Physics II:
		Nonequilibrium Statistical Mechanics} 2nd Ed. Vol.~31 (Springer-Verlag
	Berlin Heidelberg) 1991.
	
	\bibitem{batchelor1953theory}
	\Name{Batchelor G.} \Book{The Theory of Homogeneous Turbulence} Cambridge
	Science Classics (Cambridge University Press) 1953.
	
	\bibitem{Hellinger_2021b}
	\Name{Hellinger P., Papini E., Verdini A., Landi S., Franci L., Matteini L.
		\and Montagud-Camps V.} \REVIEW{Astrophys. J.}{917}{2021}{101}.
	
	\bibitem{Vembu_1961}
	\Name{Vembu S.} \REVIEW{Q. J. Math.}{12}{1961}{165}.
	
	\bibitem{LeesAluie_2019}
	\Name{Lees A. \and Aluie H.} \REVIEW{Fluids}{4}{2019}{}.
	
	\bibitem{Steele_2004}
	\Name{Steele J.~M.} \Book{The Cauchy-Schwarz Master Class: An Introduction to
		the Art of Mathematical Inequalities} (Cambridge University Press) 2004.
	
	\bibitem{Eyink_1995}
	\Name{Eyink G.~L.} \REVIEW{J. Stat. Phys.}{78}{1995}{335}.
	
	\bibitem{Eyink_1995Besov}
	\Name{Eyink G.~L.} \REVIEW{J. Stat. Phys.}{78}{1995}{353}.
	
	\bibitem{eyink_2008notes}
	\Name{Eyink G.~L.} \Book{Turbulence theory,} Course notes, The Johns Hopkins
	University (2008).
	
	\bibitem{Wang_2013}
	\Name{Wang J., Yang Y., Shi Y., Xiao Z., He X.~T. \and Chen S.} \REVIEW{Phys.
		Rev. Lett.}{110}{2013}{214505}.
	
	\bibitem{Hellinger_2021a}
	\Name{Hellinger P., Verdini A., Landi S., Papini E., Franci L. \and Matteini
		L.} \REVIEW{Phys. Rev. Fluids}{6}{2021}{044607}.
	
	\bibitem{Kida_1990}
	\Name{Kida S. \and Orszag S.~A.} \REVIEW{J. Sci. Comput.}{5}{1990}{85}.
	
	\bibitem{CHassaing1985}
	\Name{{Chassaing} P.} \REVIEW{Journal de Mecanique Theorique et
		Appliquee}{4}{1985}{375}.
	
	\bibitem{Eyink_2009}
	\Name{Eyink G.~L. \and Aluie H.} \REVIEW{Phys. Fluids}{21}{2009}{115107}.
	
	\bibitem{Capocci_2025}
	\Name{Capocci D.} \REVIEW{Phys. Rev. Fluids}{10}{2025}{L022601}.
	
	\bibitem{BUNNER_TRYGGVASON_2002}
	\Name{Bunner B. \and Tryggvason G.} \REVIEW{J. Fluid Mech.}{466}{2002}{17}.
	
	\bibitem{Pandey_Ramadugu_Perlekar_2020}
	\Name{Pandey V., Ramadugu R. \and Perlekar P.} \REVIEW{J. Fluid
		Mech.}{884}{2020}{R6}.
	
	\bibitem{Ramadugu_Pandey_Perlekar_2020}
	\Name{Ramadugu R., Pandey V. \and Perlekar P.} \REVIEW{The Eur. Phys. J.
		E}{43}{2020}{73}.
	
	\bibitem{Aniszewski2021}
	\Name{Aniszewski W., Arrufat T., Crialesi-Esposito M., Dabiri S., Fuster D.,
		Ling Y., Lu J., Malan L., Pal S., Scardovelli R., Tryggvason G., Yecko P.
		\and Zaleski S.} \REVIEW{Comput. Phys. Commun.}{263}{2021}{107849}.
	
\end{thebibliography}
\end{document}